\RequirePackage{fix-cm}
\documentclass[twocolumn,epjc3,final]{svjour3}  
\smartqed  
\RequirePackage{graphicx}
\RequirePackage{amsmath}
\RequirePackage{siunitx, tabularx, booktabs}
\RequirePackage{hyphenat}
\RequirePackage[numbers,sort&compress]{natbib}
\RequirePackage[colorlinks,citecolor=blue,urlcolor=blue,linkcolor=blue]{hyperref}
\newcommand{\figwidth}{0.49\textwidth}
\journalname{Eur. Phys. J. C}
\begin{document}

\title{Search for Neutrinos from Dark Matter Self\hyp{}Annihilations in the center of the Milky Way with 3 years of IceCube/DeepCore}

\onecolumn
\author{IceCube Collaboration: M.~G.~Aartsen\thanksref{Adelaide}
\and M.~Ackermann\thanksref{Zeuthen}
\and J.~Adams\thanksref{Christchurch}
\and J.~A.~Aguilar\thanksref{BrusselsLibre}
\and M.~Ahlers\thanksref{Copenhagen}
\and M.~Ahrens\thanksref{StockholmOKC}
\and I.~Al~Samarai\thanksref{Geneva}
\and D.~Altmann\thanksref{Erlangen}
\and K.~Andeen\thanksref{Marquette}
\and T.~Anderson\thanksref{PennPhys}
% ...
}
\authorrunning{IceCube Collaboration}
\thankstext{a}{Earthquake Research Institute, University of Tokyo, Bunkyo, Tokyo 113-0032, Japan}
\institute{
% ...
Department of Physics, University of Adelaide, Adelaide, 5005, Australia \label{Adelaide}
% ...
\and Universit\'e Libre de Bruxelles, Science Faculty CP230, B-1050 Brussels, Belgium \label{BrusselsLibre}
% ...
\and Dept.~of Physics and Astronomy, University of Canterbury, Private Bag 4800, Christchurch, New Zealand \label{Christchurch}
% ...
\and Niels Bohr Institute, University of Copenhagen, DK-2100 Copenhagen, Denmark \label{Copenhagen}
% ...
\and Erlangen Centre for Astroparticle Physics, Friedrich-Alexander-Universit\"at Erlangen-N\"urnberg, D-91058 Erlangen, Germany \label{Erlangen}
\and D\'epartement de physique nucl\'eaire et corpusculaire, Universit\'e de Gen\`eve, CH-1211 Gen\`eve, Switzerland \label{Geneva}
% ...
\and Department of Physics, Marquette University, Milwaukee, WI, 53201, USA \label{Marquette}
% ...
\and Oskar Klein Centre and Dept.~of Physics, Stockholm University, SE-10691 Stockholm, Sweden \label{StockholmOKC}
% ...
\and Dept.~of Physics, Pennsylvania State University, University Park, PA 16802, USA \label{PennPhys}
% ...
\and DESY, D-15735 Zeuthen, Germany \label{Zeuthen}
} 

\date{\today}

\maketitle
\twocolumn

\begin{abstract}
We present a search for a neutrino signal from dark matter self\hyp{}annihilations in the Milky Way using the IceCube Neutrino Observatory (IceCube).
In 1005 days of data we found no significant excess of neutrinos over the background of neutrinos produced in atmospheric air showers from cosmic ray interactions. 
We derive upper limits on the velocity averaged product of the dark matter self\hyp{}annihilation cross section and the relative velocity of the dark matter particles $\langle\sigma_{\text{A}}v\rangle$. Upper limits are set for dark matter particle candidate masses ranging from 10\,GeV up to 1\,TeV while considering annihilation through multiple channels.
This work sets the most stringent limit on a neutrino signal from dark matter with mass between 10\,GeV and 100\,GeV, with a limit of $1.18\cdot10^{-23}\text{ cm}^3\text{s}^{-1}$ for 100\,GeV dark matter particles self\hyp{}annihilating via $\tau^+\tau^-$ to neutrinos (assuming the Navarro-Frenk-White dark matter halo profile).
\end{abstract}

\section{Introduction}
\label{sec:intro}
With the increasingly strong indications of the existence of extended halos of dark matter surrounding galaxies and galaxy clusters \cite{VandenBergh1999}, there is much interest with\-in the particle physics community to determine the nature and properties of dark matter \cite{bertone}.
The frequently considered hypothesis is that dark matter consists of stable massive particles interacting feebly with Standard Model particles. 
The density of dark matter particles today is determined by the `freeze-out' \cite{zeldovich,vysotsky,lee,wolfram} in the early universe when the thermal equilibrium can no longer be sustained as the universe expands and cools down.
This work focuses on a generic candidate particle for dark matter referred to as a Weakly Interacting Massive Particle (WIMP) \cite{gunn,srednicki,rodaz,bengtsson}, though this search is sensitive to any self\hyp{}annihilating dark matter particle with a coupling to the Standard Model resulting in a flux of neutrinos.
The source considered is the Milky Way galaxy, which is embedded in a spherical halo of dark matter \cite{einasto65,einasto68,Navarro:1996ce,Burkert:1995jr,moore}.
For a given halo density profile, the total amount of dark matter in the line of sight from Earth can be determined \cite{bergstrom}.

If WIMPs can self\hyp{}annihilate into Standard Model particles and the dark matter density is sufficiently high, an excess of neutrinos and photons should be observed from parts of the sky with a large amount of dark matter, above the background of muons and neutrinos produced in the Earth's atmosphere.
Although photons produced in such annihilations are far easier to detect, it is still of interest to consider scenarios where only neutrinos are produced \cite{Yuksel:2007ac}.

The targeted neutrino signal is estimated from a dataset of simulated neutrino events reweighted to the energy and directional distribution of dark matter in the Milky Way.
The background is uniform in right ascension and is estimated from experimental data.
A shape likelihood analysis on the reconstructed neutrino direction is used to estimate the fraction of events possibly originating from the targeted signal.
From the signal fraction a limit on the signal flux is calculated and the corresponding value of $\langle\sigma_\text{A} v\rangle$ can be determined for any combination of WIMP mass and WIMP annihilation channel to neutrinos.

This search focuses on charged-current muon neutrinos because their directions can be accurately reconstructed.
However, other neutrino flavors and events from neutral-current neutrino interaction are also present in the final selection (ensuring the most inclusive limits).

\section{IceCube Neutrino Observatory}
\label{sec:icecube}
\begin{figure}
  \includegraphics[width=\figwidth]{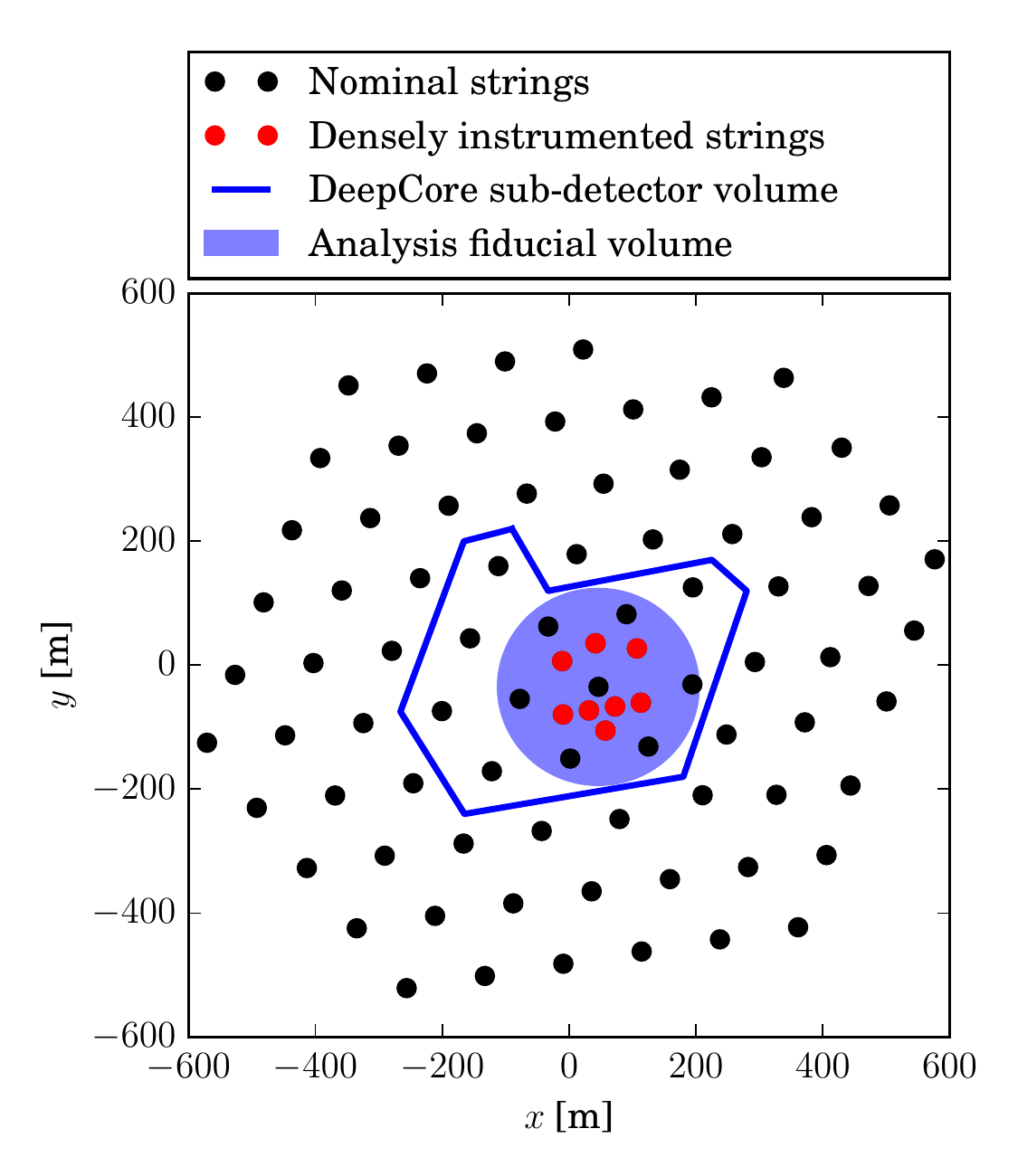}
\caption{The horizontal position of the deployed strings in the IceCube coordinate system.
The blue line shows the strings constituting the DeepCore subdetector, strings outside of this region are used in the initial event rejection.
The fiducial volume used in the final analysis is indicated with the solid blue region consisting of both nominal and dense strings.}
\label{fig:geo}
\end{figure}

IceCube detects Cherenkov light from charged particles moving through one cubic kilometer of very transparent ice underneath the South Pole \cite{1748-0221-12-03-P03012,Achterberg:2006kh}.
The array consists of 78 vertical strings in a hexagonal grid with 60 digital optical modules (DOMs) \cite{Abbasi:2010kx} spaced evenly on each string every 17\,m between 1450-2450\,m below the surface.
The spacing between these nominal strings is approximately 125\,m (as shown by the black dots in Figure \ref{fig:geo}).
In addition there are eight strings in the central area (red dots in Figure \ref{fig:geo}) with the DOMs more densely spaced constituting the infill IceCube\slash DeepCore \cite{Abbasi:2012ho}.

The fiducial volume used in this work is defined by DOMs located 2140-2420\,m below the surface situated on the most central strings (indicated with a solid blue region in Figure \ref{fig:geo}).
The rest of IceCube is used as a veto volume to reject incoming and through-going atmospheric muons.

The strings outside the DeepCore sub\hyp{}detector volume (indicated with a blue line in Figure \ref{fig:geo}) are only used in the initial filtering of triggered data, and are chosen to be shielded by three rows of DOMs from the edge of the array.

\section{Signal expectation}
\label{sec:sim}
For WIMPs self\hyp{}annihilating to various Standard Model particles (leptons, quarks, or bosons), the decay chain of the particles will ultimately produce leptons and photons.
Depending on the WIMP mass ($m_{\text{DM}}$) and annihilation channel, a number of neutrinos will be produced in the decay chain, propagate to Earth, and can be detected in neutrino observatories.

\begin{figure}
  \includegraphics[width=\figwidth]{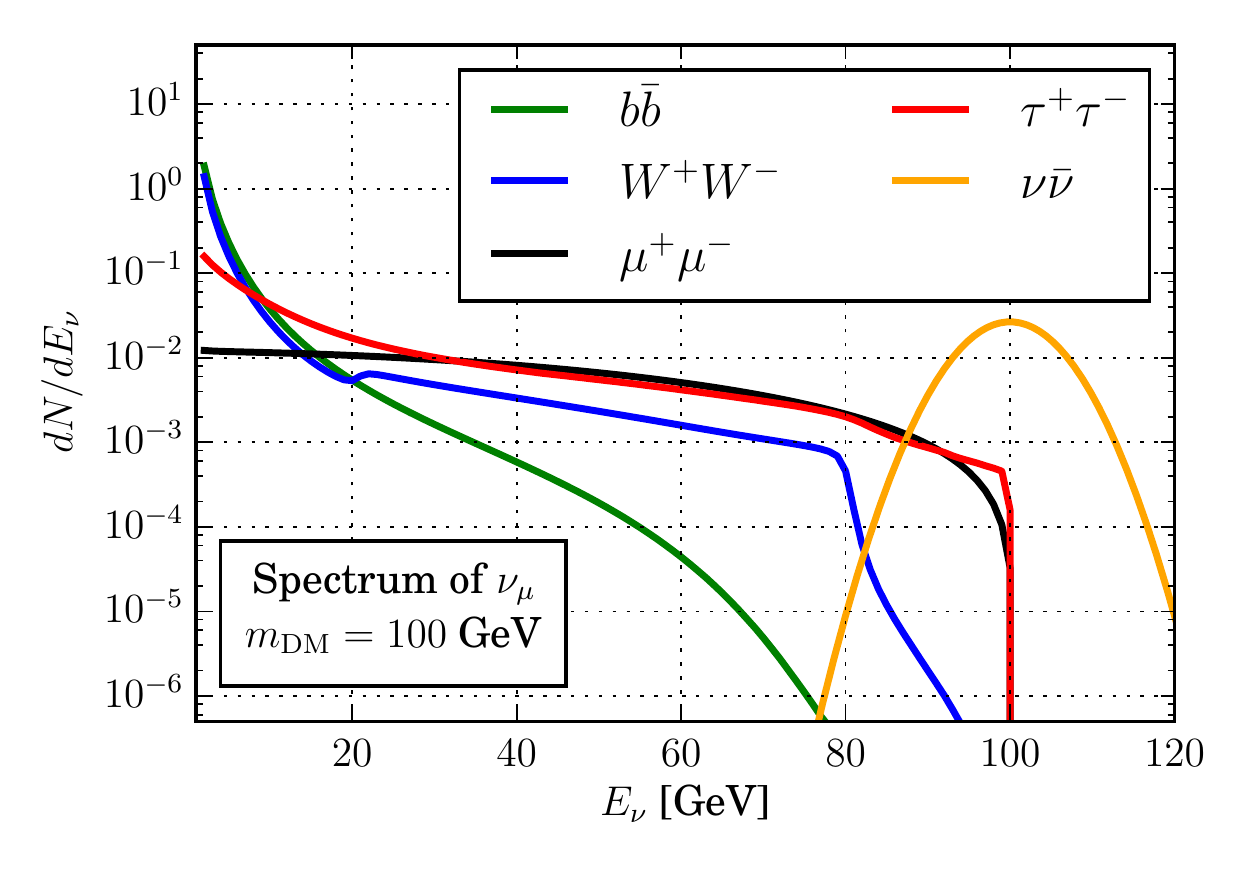}
\caption{Energy spectrum of muon neutrinos at Earth produced in the annihilation and subsequent decay of various Standard Model particles created in the annihilation of a 100\,GeV WIMP. The line spectrum of the $\nu\bar{\nu}$-channel is modeled by a Gaussian with a width of 5\% of $m_{\text{DM}}$.}
\label{fig:dnde}
\end{figure}

Using \texttt{PYTHIA} \cite{pythia6,pythia8}, a generic resonance with twice the WIMP mass is forced to decay through one of the particle pairs (annihilation channels) considered and the energy spectra of the resulting neutrinos are recorded for all three neutrino flavors.
This work considers WIMPs with masses from 10-1000\,GeV self\hyp{}annihilating through either $b$\hyp{}quarks ($b\bar{b}$), $W$\hyp{}bosons ($W^+W^-$), muons ($\mu^+\mu^-$), or taus ($\tau^+\tau^-$) to neutrinos.
Annihilation directly to neutrinos ($\nu\bar{\nu}$) is also considered.
In Figure \ref{fig:dnde} the energy spectrum, $dN/dE$, of muon neutrinos from a pair of 100\,GeV WIMPs is presented for the annihilation channels considered in this analysis.
The energy spectrum is shown after applying long baseline oscillations (determined from parameters in \cite{Forero:2014fb}).

For the $W^+W^-$-channel only WIMP masses above the mass of the $W$ boson are probed.
The energy spectrum of the $\nu\bar{\nu}$-channel is dominated by the line at $m_{\text{DM}}$, which is modeled with a Gaussian distribution with a width of 5\% of $m_{\text{DM}}$.
This width provides the possibility to use the same simulated dataset, while still being consistent with a line spectrum after smearing by the event reconstruction.
For the signal from the $\nu\bar{\nu}$-channel a flavor ratio produced at the source of $(\nu_e:\nu_{\mu}:\nu_{\tau})=(1:1:1)$ is used (though the most conservative limits are found for a flavor ratio of $(1:0:0)$ at source resulting in 10-15\% weaker limits).
The results will be presented with a 100\% branching ratio for each annihilation channel considered.

The rate of WIMP self\hyp{}annihilation seen in a given solid angle is determined from the integrated dark matter density along the line of sight (los) through the dark matter halo in the Milky Way.
Although there remain uncertainties about the dark matter density profile \cite{Diemand:2011kh}, a spherical profile is assumed with one of two standard radial distributions: 
Navarro-Frenk-White (NFW) \cite{Navarro:1996ce} and Burkert \cite{Burkert:1995jr} with parameter values from \cite{Nesti:2013wm}.
The resulting rate of dark matter self\hyp{}annihilations along the line of sight is strongly dependent on the assumed halo density, with the largest discrepancies near the center of the Milky way where the density is largest.
Because of the large uncertainty on the model parameters the dark matter halo model constitutes the largest systematic uncertainty.

The resulting differential flux of \textit{signal neutrinos} produced by WIMP self\hyp{}annihilation in the dark matter halo of the Milky Way from a solid angle of the sky, $\Delta\Omega$, is given as
\begin{align}
\frac{d\Phi}{dE}(\Delta\Omega) = \frac{\langle\sigma_{\text{A}}v\rangle}{4\pi\cdot2m_{\text{DM}}^2}\frac{dN}{dE}\int_{\text{los}}\rho^2(r(l,\Delta\Omega))dl,\label{eq:flux}
\end{align}
where the $4\pi$ arises from a spherically symmetric annihilation, $l$ is the line of sight through the dark matter halo with density profile $\rho(r)$ as a function of radius $r$, and the factor of $1/2$ and the squared WIMP mass and halo density profile arise from the fact that two WIMPs are needed in order to annihilate. 

A sample of neutrino events of each flavor is generated with energies between 1-1000\,GeV using \texttt{GENIE} \cite{genie} and weighted to the targeted flux of Equation \eqref{eq:flux} according to their flavor, energy, and arrival direction for each combination of $m_{\text{DM}}$, annihilation channel and dark matter halo density profile.
This neutrino sample provides the distribution of the targeted signal that is used in the shape likelihood analysis to determine the fraction of possible signal events in the experimental data.

\section{Background estimation}
The background consists of neutrinos with other astrophysical origin, atmospheric neutrinos, and atmospheric muons.
At the energies considered, the event sample is dominated by atmospheric neutrinos and muons produced in cosmic ray induced air showers.
The cosmic ray flux is isotropic in right ascension, so the atmospheric background can be estimated from experimental data by randomizing the arrival times of each event.
Since IceCube has a uniform exposure this corresponds to randomizing the right ascension values, which has shown in a previous analysis to be an unbiased approach to estimate the background \cite{Aartsen:2015fa}.

The largest expected background contribution is from down-going atmospheric muons.
This is because IceCube is located at the South Pole, so the center of the Milky Way (corresponding to the direction with the strongest signal) will be above the horizon, where there will also be the highest rate from atmospheric muons.
Therefore the goal of the initial event selection is to reduce the rate of atmospheric muons.
The overall analysis is verified using a simulation of atmospheric muons generated with \texttt{CORSIKA} \cite{corsika} compared to the experimental data.
The rate of simulated background is within 5\% of the experimental data (see Table \ref{tab:rates}). 

The other significant background contribution is atmospheric neutrinos.
They arrive at IceCube from all directions and cannot be distinguished from extraterrestrial neutrinos event-by-event.
However, from the full statistical ensemble the distributions can be distinguished by their energy and arrival direction.
Simulated \texttt{GENIE} neutrino datasets are used for estimating the fraction of atmospheric neutrinos in the final selection of the experimental data, using the atmospheric neutrino flux model described in \cite{Honda:2015jn}.
The simulated atmospheric neutrinos do not impact the result, as the combined background is estimated from experimental data.

The extra-galactic neutrino background can be distinguished from the WIMP neutrino signal by the arrival distribution, which is not necessarily the case for galactic neutrinos.
But at the energies considered, both are expected to be more than three orders of magnitude below the background of atmospheric neutrinos.

\section{Event selection}
\label{sec:event}
The event selection was optimized for the signal of muon neutrinos from 100\,GeV WIMPs self\hyp{}annihilating through the $W^+W^-$\hyp{}channel (benchmark channel) and is applied event wise on the experimental data and the simulated event samples.
The aim is to select high quality neutrino induced muons, signified by elongated event topologies (referred to as \textit{tracks}) starting inside IceCube/DeepCore.

The neutrino induced muons need to be distinguished from the muons produced in the atmosphere.
All atmospheric muons detected in IceCube penetrate through the veto volume. 
The corresponding hits (reconstructed pulses from one or more detected photons) can therefore be used to identify and remove these through-going tracks.

The event selection is a multi-step background rejection procedure that reduces the atmospheric muons by seven orders of magnitude.

The first step is to clean the DOM hits to remove noise so that the precision of the reconstruction is not degraded.
Next, events with more than one hit in the volume outside the DeepCore sub-detector volume causally connected to a charge weighted center of gravity in the fiducial volume within a predefined time window and distance are removed.
This filters out atmospheric muons with very basic event information.

By requiring more than ten hits distributed on at least four strings nearly all noise-only events are removed.
In addition, this requirement ensures that the events can be well reconstructed.
The three first hits in the event are required to be in the fiducial volume, as that is more likely to indicate a starting event and thus reduce the rate of penetrating atmospheric muons.
The events are reconstructed to preliminarily estimate the direction and interaction point of the candidate neutrino-induced muon.
The events with a preliminary zenith angle for the arrival direction of $\text{zen} > \text{zen}_{\text{GC}}+20^{\circ}$ or  $\text{zen} < \text{zen}_{\text{GC}}-10^{\circ}$ are rejected, where $\text{zen}_{\text{GC}}$ denotes the zenith of the Galactic center.
The cut is asymmetric because the atmospheric muon background is increasingly larger towards a zenith of zero (i.e. the southern celestial pole).
A containment cut is used to keep only events that have a reconstructed interaction vertex within a cylinder with a radius corresponding to the analysis volume depicted on Figure \ref{fig:geo}.
In addition cuts are applied on track quality \cite{Neunhoffer:2006bs}.

By considering the hits in the veto volume that are cleaned away (as possible noise), clusters are determined for hits that are within 250\,m and 1000\,ns from each other and are registered earlier than the first quantile of cleaned hits. 
These clusters are required to have fewer than three hits, as larger clusters are generally observed more often for penetrating atmospheric muons.

A cone with a 20 degree opening angle aimed towards the arrival direction is used to check for hits in the uncleaned hit series within 1\,$\mu$s of the interaction.
At most one hit is allowed, since events starting within the fiducial volume should have zero hits within the cone, but one accidental noise hit is allowed.
Due to the high rate of atmospheric muons versus possible signal neutrinos, there is a class of background muon events where sparse hits in the veto volume are removed during the hit cleaning.
The uncleaned hits in a cylinder with a radius of 250\,m pointed towards the arrival direction starting behind the interaction vertex, are used to calculate the likelihood value for the reconstructed track.
A high likelihood value indicates that the track probably originated from a penetrating muon, for which the hits deposited in the veto volume are erroneously cleaned away. 

At the energies considered in this analysis, the reconstruction must take into account both the ha\-dron\-ic cascade and the muon produced in a typical muon neutrino charged current interaction.
With the experimental data event rate reduced by six orders of magnitude from 2\,kHz to 3.7\,mHz by the cuts described, a more specific event reconstruction can be run.
This low energy specialized event reconstruction fits all relevant parameters (direction, interaction vertex, muon track length, and ha\-dron\-ic cascade energy) simultaneously and takes into account both DOMs that did and did not detect any light.
In order to thoroughly sample the complex likelihood space of the full 8-dimensional parameter space the Bayesian sampling inference tool Multi\-Nest \cite{Feroz:2009eba} is used. 

The final step of the event selection is a multivariate analysis using a Boosted Decision Tree (BDT) \cite{Freund:1997ft}.
First of all, the direction and vertex information from the specialised event reconstruction are used along with the number of hits in a 10 degree opening angle veto cone, updated with the specialised event reconstruction.
Further, the difference in likelihood in reconstructing the event with a finite track (expected from a neutrino induced starting muon) compared to an infinite track (expected for a through-going atmospheric muon) is used.
An additional veto technique traces back in the direction of arrival from the interaction vertex to look for charge on DOMs that would identify the event as a through-going muon misidentified as a starting event.
Both the number of hits and the total charge identified by the veto are used in the BDT.

The events are selected based on the BDT score, optimized for the best sensitivity to the benchmark signal of a 100\,GeV WIMP annihilating through $W^+W^-$.
The same cut value is used across multiple WIMP masses and annihilation channels.

\begin{figure}
  \includegraphics[width=\figwidth]{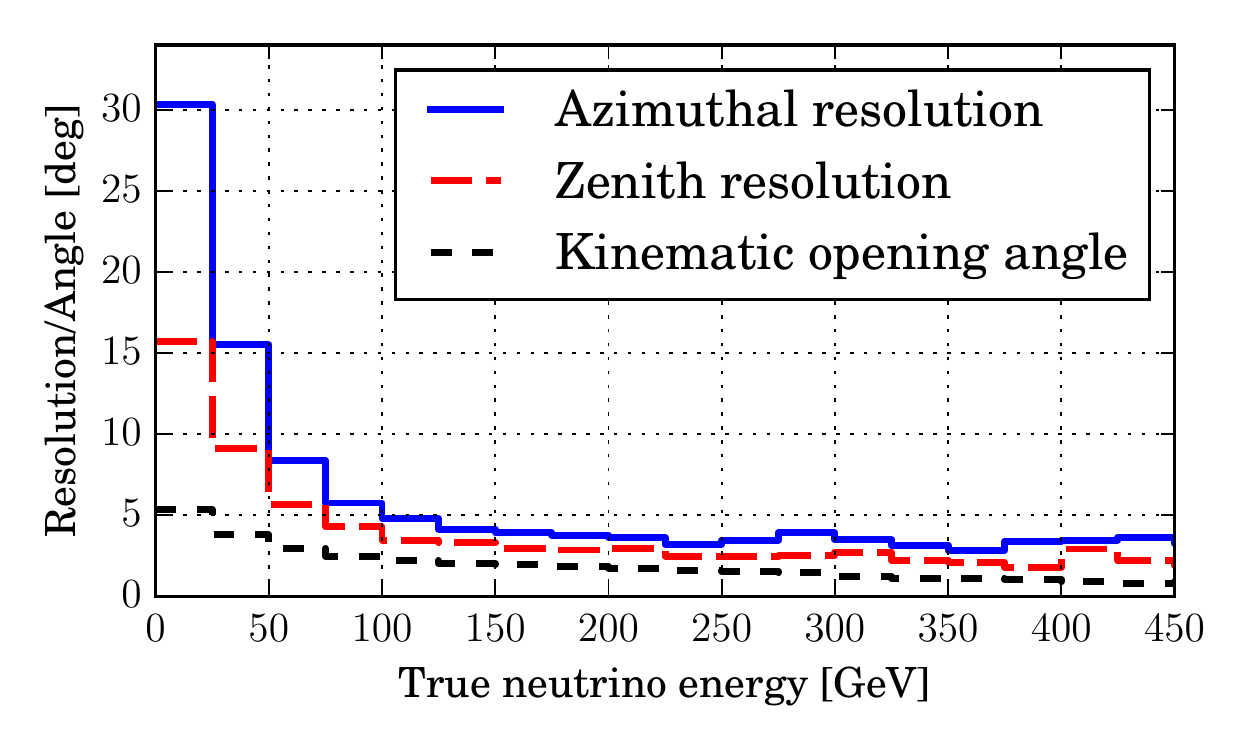}
\caption{Resolution of the azimuthal and zenith direction of $\nu_{\mu}$ in the event sample, shown as a function of energy, compared to the kinematic opening angle.}
\label{fig:resen}
\end{figure}

\begin{figure}
  \includegraphics[width=\figwidth]{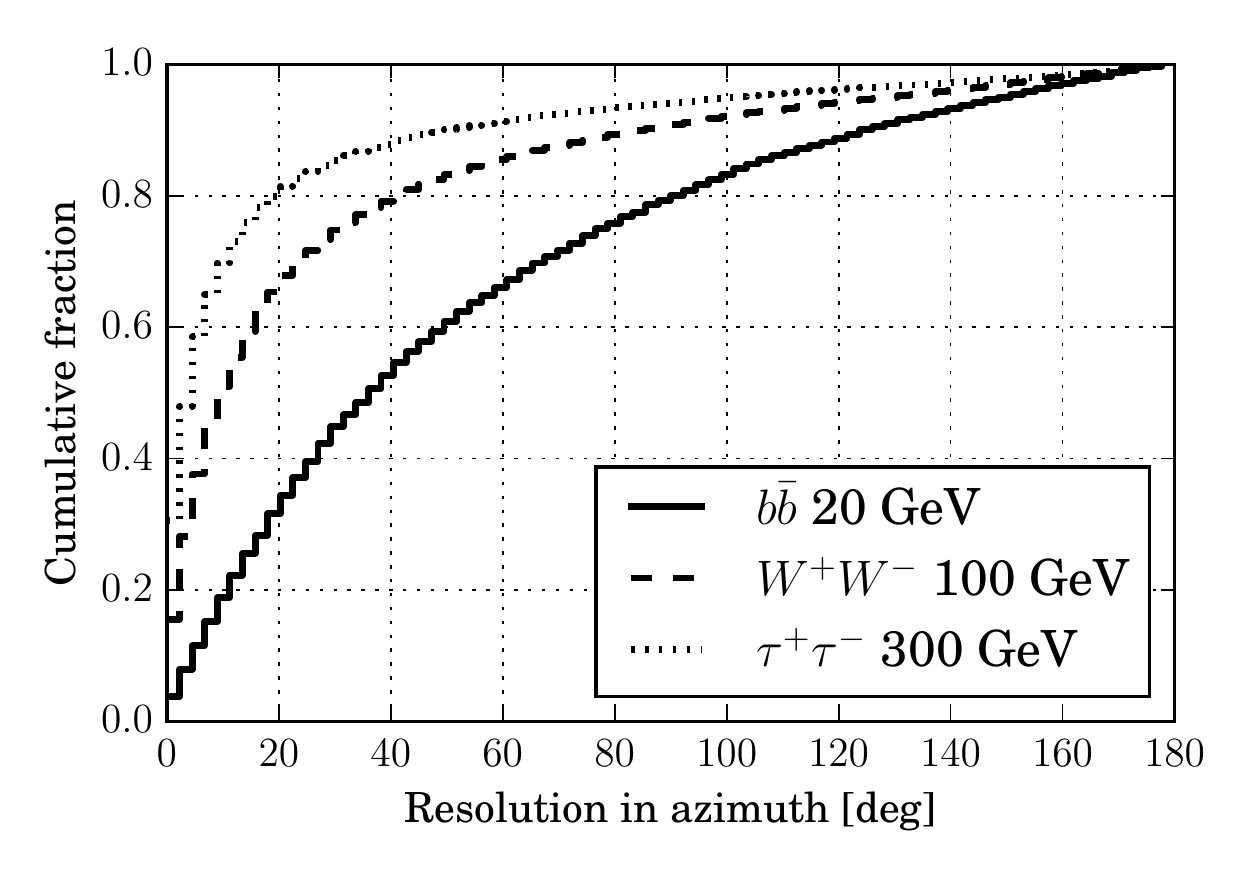}
\caption{Cumulative distribution of the resolution of the azimuthal direction of $\nu_{\mu}$ in the final event sample, for various WIMP masses and annihilation channels.}
\label{fig:rescumu}
\end{figure}

The median resolution in azimuthal angle is presented in Figure \ref{fig:resen} as a function of true neutrino energy. 
Because the azimuthal angle maps directly to right ascension, it provides the dominating separation between signal and background.
A comparison of three combinations of WIMP mass and annihilation channel is presented in Figure~\ref{fig:rescumu}, illustrating a better resolution for cases where the neutrino spectrum continues to higher energies.


\begin{table*}[t]
\caption{Event rates for the various components expected in the experimental data given in mHz, and the signal neutrinos are presented as percentage of the events at filtered level for the benchmark signal (annihilation of a 100\,GeV WIMP to $W^+W^-$).
Everything but the experimental data is based on simulation.
The atmospheric muons rates are based on the GaisserH3a energy spectrum \cite{Gaisser:2012em}.
The atmospheric neutrinos rates are based on neutrino oscillation parameters in \cite{GonzalezGarcia:2014di}. 
Due to vanishing rates at higher levels the rate of atmospheric $\nu_{\tau}$ are not listed.
}
\label{tab:rates}
\centering
\begin{tabularx}{0.99\textwidth}{l c S[table-format=3.3] S[table-format=2.3] *{2}{S[table-format=1.3]}}
\toprule
{Dataset} & {DeepCore filtered trigger data} & {Quality cuts} & {Atm. bkgd. rejection} & {Pre-BDT linear cuts} & {BDT}\\
\midrule
Experimental data  & {$\sim15\cdot10^3$} & 655.0 & 36.73 & 3.59 & 0.27 \\
Atmos. $\mu$ (H3a) & {$\sim9.5\cdot10^3$}  & 656.9 & 37.88 & 3.53 & 0.19 \\
Atmos. $\nu_{\mu}$ & 6.49 & 2.14 & 0.319 & 0.199 & 0.07 \\ 
Atmos. $\nu_{e}$   & 2.06 & 0.43 & 0.043 & 0.027 & 0.01 \\
Noise-only events  &  {$\sim6.6\cdot10^3$} & 0.1 & 0 & 0 & 0 \\
\midrule
Signal $\nu_{\mu}$ & {100\%} & {70.48\%} & {14.67\%} & {9.29\%} & {6.20\%} \\
Signal $\nu_e$     & {100\%} & {81.31\%} & {10.94\%} & {6.94\%} & {4.96\%} \\
Signal $\nu_{\tau}$& {100\%} & {80.61\%} & {10.63\%} & {7.29\%} & {5.88\%} \\
\bottomrule
\end{tabularx}
\end{table*}

The final event selection results in a data rate of 0.27\,mHz, corresponding to a reduction by 7 orders of magnitude from the initial triggering of the data, while retaining 6\% of the benchmark signal of muon neutrinos.
No cuts have been incorporated to explicitly remove non-muon neutrino flavors. 
In the final event sample the non-muon neutrinos of the targeted signal are present with a combined rate comparable to that of muon neutrinos.
Using the \texttt{GENIE} neutrino simulation weighted to the atmospheric flux model, it is estimated that atmospheric neutrinos constitute one quarter of the final experimental data.
A summary of the event selection rates and signal efficiency is given in Table \ref{tab:rates}.

\begin{figure}
  \includegraphics[width=\figwidth]{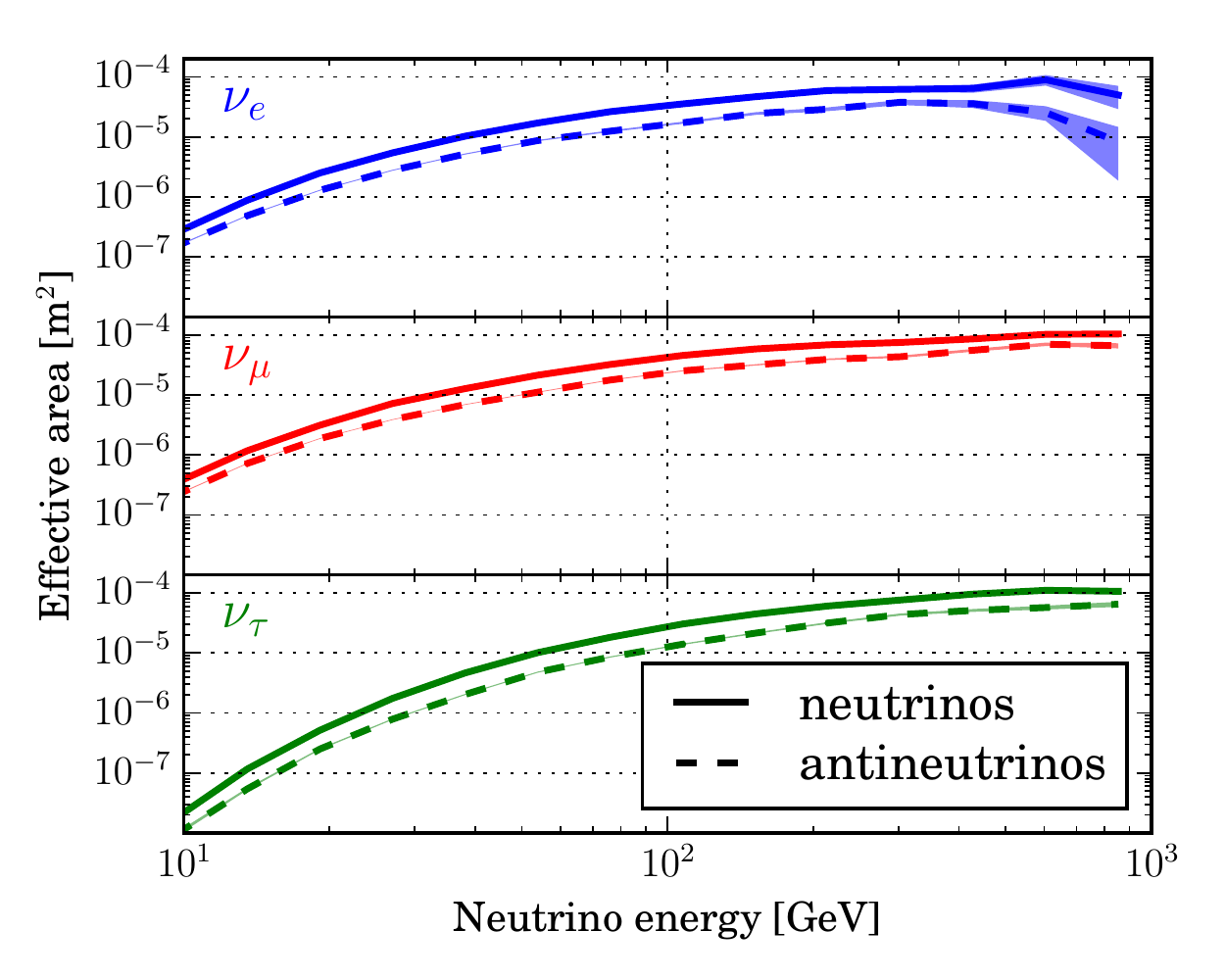}
\caption{Effective area of final event sample for the three neutrino flavors with both charged- and neutral-current interactions combined.}
\label{fig:aeff}
\end{figure}

In Figure \ref{fig:aeff} the effective area at the final level is presented for the individual neutrino flavors combining both neutral- and charged-current neutrino interactions.

\section{Analysis method}
\label{sec:ana}
The final event sample is filled into 2D histograms with bins covering the range $[0,2\pi]$ rad in right ascension (RA) and $[-1,1]$ rad in declination (Dec) using the reconstructed values from the specialised event reconstruction.
The bin width is chosen to be 0.4 and 0.63 radians for RA and declination, respectively, based on the resolution of the event reconstruction.
In order to ensure a consistent analysis the same bin width is chosen for the combination of WIMP mass and annihilation channel that exhibits the worst resolution.
The 2D distributions constitute the probability density functions (PDFs) used in the shape likelihood analysis described below.
The shape of the 2D distribution of experimental data produces the \textit{data PDF} which is compared to the expectation from the weighted signal distributions (or \textit{signal PDF}) and the estimated background distribution which is constructed from the experimental data.

The experimental data scrambled in RA (assigned a random RA value for each event) consist of a component of scrambled background and potential signal (also scrambled):
\begin{align}
\text{PDF}_{\text{scr. data}} = (1-\mu)\text{PDF}_{\text{scr. bkg}} + \mu\text{PDF}_{\text{scr. sig}}, \label{eq:scrdata}
\end{align}
where $\mu\in[0,1]$ parametrizes the fraction of signal in the total sample.

From equation \ref{eq:scrdata} the \textit{background PDF} can be estimated from the experimental data (by subtracting the scrambled signal) under the hypothesis that the background is uniform in RA and hence invariant under scrambling.

The total fraction of events within a specific bin  $i\in[\text{bin}_{\text{min}},\text{bin}_{\text{max}}]$ is calculated as a function of the signal fraction as  
\begin{align}
f(i|\mu) = \mu\text{PDF}_{\text{sig}}(i) + (1-\mu)\text{PDF}_{\text{scr. bkg.}}(i). \label{eq:fraction}
\end{align}

\begin{figure}
  \includegraphics[width=\figwidth]{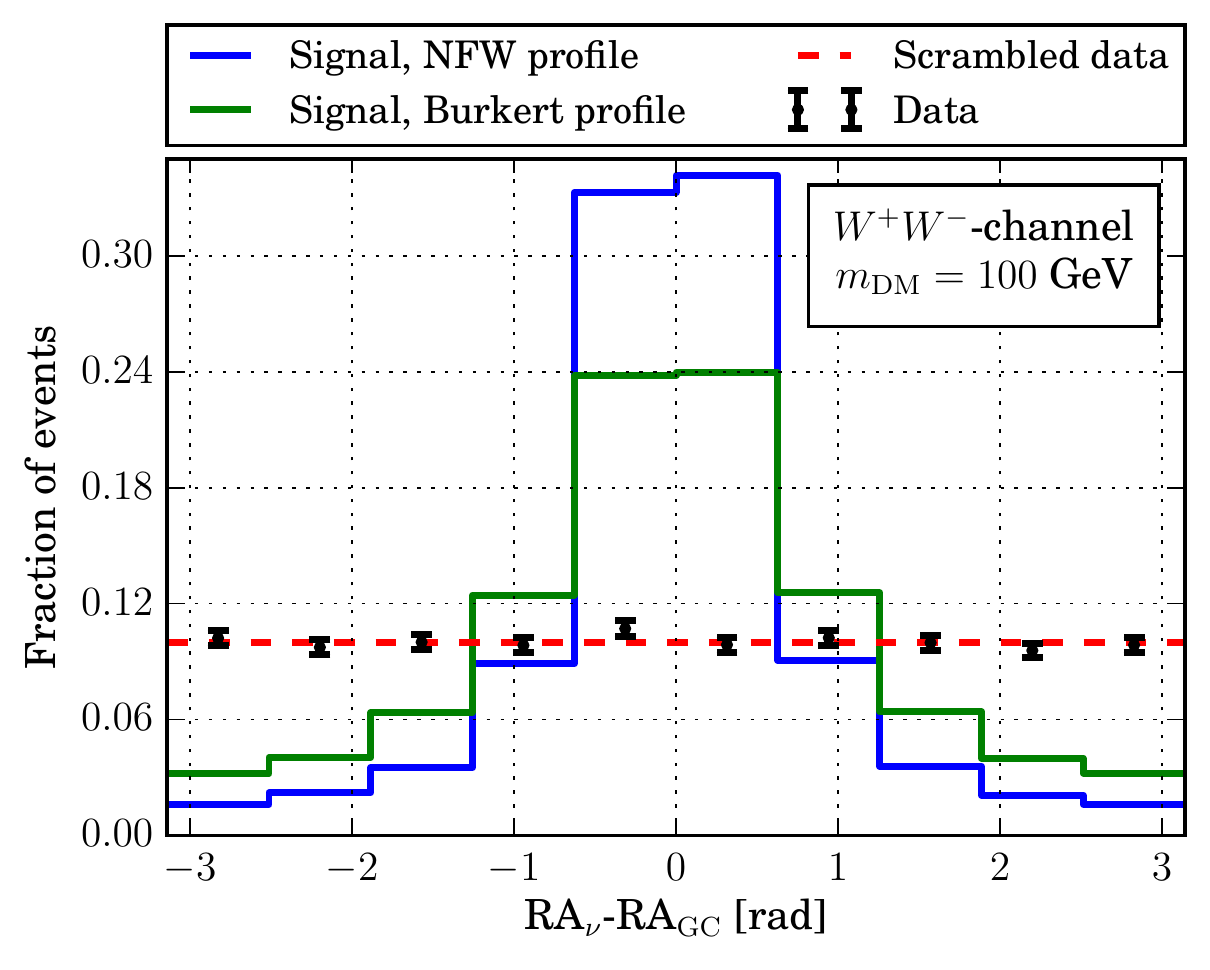}
\caption{Event distribution in right ascension (RA) relative to the galactic center (GC) of data, scrambled signal, and targeted signal for a 100\,GeV WIMP annihilation to neutrinos through the $W^+W^-$-channel (shown for a single declination bin).}
\label{fig:pdfs}
\end{figure}

In Figure \ref{fig:pdfs} an example of the relevant PDFs is presented over the full range in right ascension for a single bin in declination ($\text{dec}\in[-1/3,-2/3]$) where the largest difference between signal and background is expected.
Since the background is uniform in right ascension and the signal is peaked around the position of the center of the Milky Way, it is in right ascension that the difference between signal and background can be found.
Figure \ref{fig:pdfs} also illustrates the difference in the targeted signal between the NFW and Burkert models of the dark matter halo density profile.

With a 2D binned shape likelihood analysis, the data PDF is compared to the expectation from the background PDF and the signal PDF, for multiple combinations of WIMP mass, annihilation channel, and halo profile.
This way the most probable signal fraction is determined from the experimental data.
The likelihood is calculated by comparing the number of observed events in the individual bins $n_{\text{obs}}(i)$, assuming a Poisson uncertainty on the number of events expected, determined from the total number of events filled in the histogram $n_{\text{obs}}^{\text{total}}$ and $f(i|\mu)$ calculated in equation \ref{eq:fraction}.
This results in the following formulation of the likelihood function $\mathcal{L}(\mu)$:
\begin{align}
\mathcal{L}(\mu)&=\prod_{i=\text{bin}_{\text{min}}}^{\text{bin}_{\text{max}}}\text{Poisson}\left(n_{\text{obs}}(i)\big|n_{\text{obs}}^{\text{total}}f(i|\mu)\right).
\end{align}

Using the likelihood analysis, the best estimate of the signal fraction can be found by minimizing $-\log\mathcal{L}$, and if it is consistent with zero the 90\% confidence interval is determined applying the Feldman-Cousins approach \cite{Feldman:1998bu} to estimate the upper limit on the signal fraction $\mu_{90\%}$.
Using the simulated signal neutrinos the signal fraction can be related to $\langle\sigma_\text{A} v\rangle$.
The expected limit on $\langle\sigma_\text{A} v \rangle$ in the absence of signal is calculated from 10000 pseudo experiments sampled from the background-only PDF, from which the median value of the resulting 90\% upper limits is quoted as the \textit{sensitivity}.

\section{Systematic uncertainties}
\label{sec:syst}
The statistical uncertainty due to the limited number of events in the simulated datasets is insignificant compared to the systematic uncertainties, as the simulation holds 20 times more events than in the experimental data, after cuts.
However, all systematic uncertainties are effectively negligible compared to the astrophysical uncertainties associated with the parameters of the dark matter halo models.

The biggest systematic uncertainty arises from the modelling of the ice properties and the uncertainty on the optical efficiency of the DOMs, which increase with lower neutrino energies, and therefore for lower WIMP masses.
The precision of the detector geometry and timing are so high that the associated systematic uncertainty is negligible and therefore not included in this study.

The effect of experimental systematic uncertainties on the final sensitivity is estimated using Monte Carlo simulations of neutrinos with uncertainty values varied by $\pm 1\sigma$ from the values used in the baseline sets.
Each of the datasets with variations is run through the event selection and analysis, providing a different value for the sensitivity on $\langle\sigma_\text{A} v\rangle$.
The difference between the baseline and the variation will be quoted as the systematic uncertainty on $\langle\sigma_\text{A} v\rangle$, for each of the variations.
The systematic uncertainties are dependent on the neutrino energy, and hence on the targeted WIMP mass.
Since the background is estimated from experimental data, the variations are applied to the signal simulation only.

The optical properties of the ice in IceCube have been modelled and show an absorption and scattering length that vary with depth, generally becoming more clear in the deeper regions of IceCube.
For the experimental data there will always be a discrepancy between the ice the photons are propagating through, and the ice \cite{spice} assumed in the reconstruction (as the complicated structure of the real ice can not be perfectly modeled).
This is also the case in simulation, where the latest iteration of the ice model is used in the Monte Carlo event simulation, but because of its complexity, cannot currently be used for reconstruction.
While estimating the impact of using a different ice model for event reconstruction than used in the photon propagation simulation, it additionally accounts for the fact that the ice model in simulation is different from that used in simulation.
The effect is calculated using a variant Monte Carlo simulation with a different ice model used for the photon propagation (the same as used in the event reconstruction).
This results in a 5-15\% (depending on WIMP mass, 10\% for the benchmark channel) improvement in sensitivity on $\langle\sigma_\text{A} v\rangle$, compared to the baseline simulation.

The ice in the drill hole columns has different optical properties from the bulk ice.
The scattering length is greatly reduced due to the presence of impurities.
One effect of this column is to increase the detection probability for down\hyp{}going photons.
Since the DOMs are facing downwards, no down\hyp{}going photons would be observed without scattering.

The column ice is treated as having a much shorter geometrical scattering length: 50\,cm as a baseline \cite{spice}, implemented in simulation as photons approach the DOMs.
The uncertainty on the scattering length is covered by including variations of 30 cm and 100 cm.
This variation results in a 25-30\% reduction or 5-10\% improvement of the sensitivity on $\langle\sigma_\text{A} v\rangle$ respectively (depending on WIMP mass, 25\% and 8\% for the benchmark channel).

The photon detection efficiency of the DOMs (combining the effect of the quantum efficiency of the PMT, photon absorption by the cables in the ice, and other subdominant hardware elements) is determined to 10\% accuracy.
Increasing or decreasing the DOM efficiency in the simulation corresponds to a 5-40\% (depending on WIMP mass, 15\% for the benchmark channel) effect that symmetrically improves or reduces the sensitivity on $\langle\sigma_\text{A} v\rangle$.

The systematic uncertainties are considered to be independent and the $\pm$variation that results in the largest uncertainty for each systematic uncertainty is added in quadrature to form the total systematic uncertainty.
These are included in the final result by scaling up the limits with the total systematic uncertainty.

The dominant theoretical systematic uncertainty is related to fitted parameters of the dark matter halo profiles.
Considering the 1$\sigma$ variation on both parameters for the individual models result in a 150-200\% uncertainty on the sensitivity on $\langle\sigma_\text{A} v\rangle$.
Since this effect is theory-dependent, and may change as dark matter halo models evolve, it is not included in the total systematic uncertainty.
Instead, the results are presented for both dark matter halo models.

\section{Results}
\label{sec:result}
\begin{figure*}
  \includegraphics[width=\figwidth]{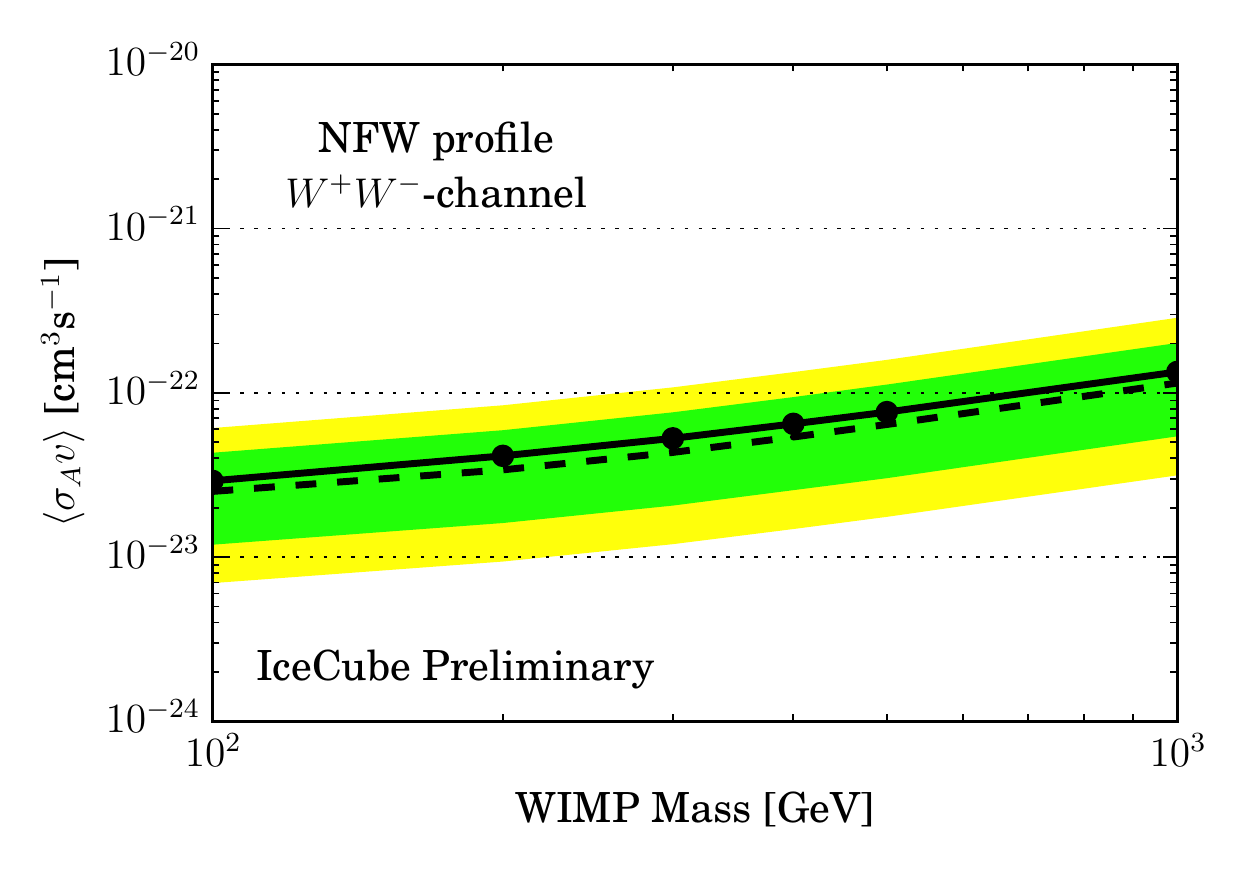}
  \includegraphics[width=\figwidth]{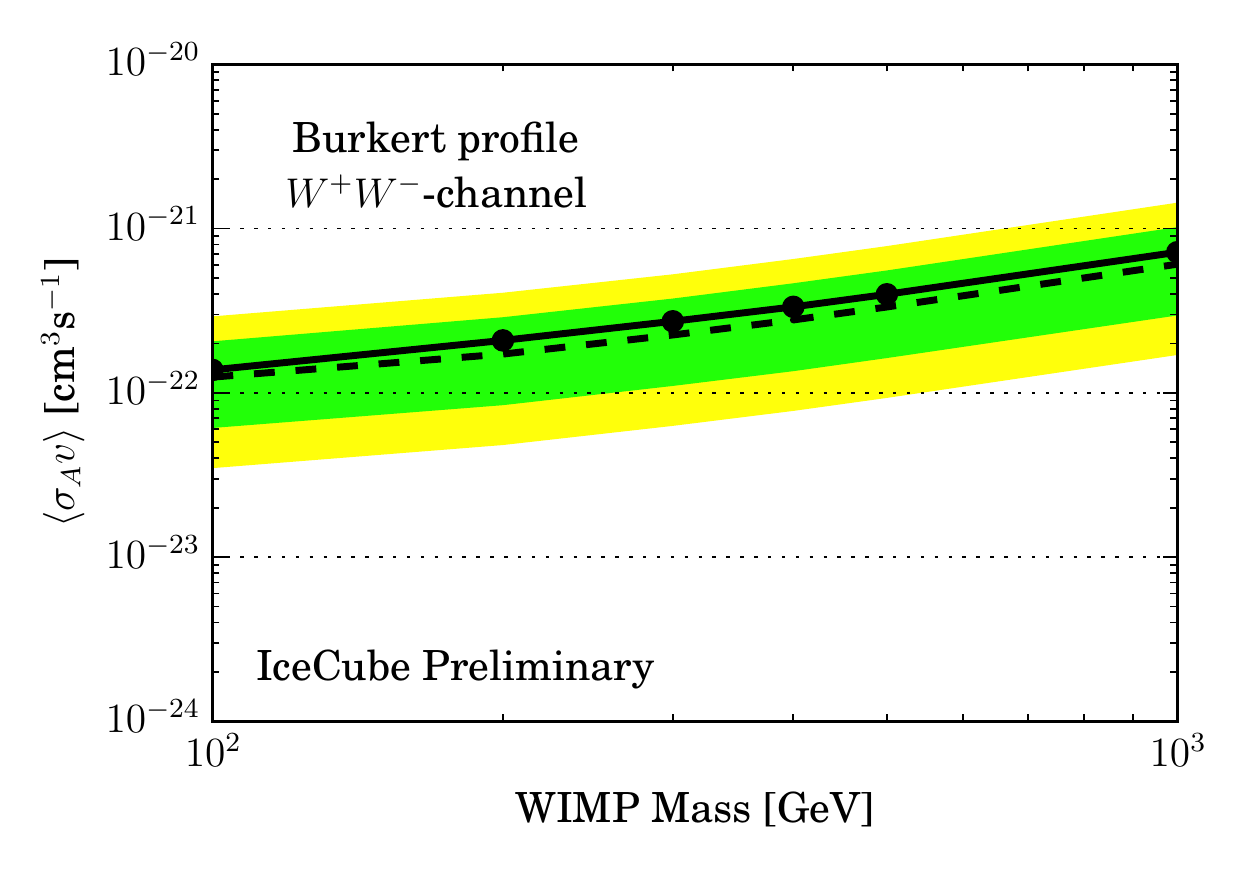}
\caption{The final limits without systematic uncertainties (solid line), compared to the sensitivity (dashed line). 
Showing the 1$\sigma$ (green band) and 2$\sigma$ (yellow band) statistical uncertainty for dark matter self\hyp{}annihilating through the $W^+W^-$ channel to neutrinos assuming a NFW (Burkert) halo profile on the left (right) plot.}
\label{fig:sens}
\end{figure*}

After the final event selection, $22\,632$ events were observed in 1005 days of IceCube data.
The data are presented in Figure \ref{fig:pdfs} illustrating that the data are compatible with the background-only hypothesis.
Since no significant excess has been observed, an upper limit on $\langle\sigma_\text{A} v\rangle$ is determined.
Figure \ref{fig:sens} shows the 90\% confidence upper limits (solid black line) for the $W^+W^-$-annihilation channel for the two dark matter halo profiles.
The colored bands represent the range of expected outcomes of this measurement with no signal present.
The result is very near the median sensitivity, and thus compatible with the background-only hypothesis, which is the case across all annihilation channels.

Tables \ref{tab:bur} and \ref{tab:nfw} show the final upper limits on $\langle\sigma_{\text{A}}v\rangle$ for all annihilation channels and WIMP masses considered in this analysis after accounting for the systematic uncertainties.

\begin{table}
\caption{Upper limits on the self\hyp{}annihilation cross section assuming the NFW halo profile.}
\label{tab:bur}
\begin{tabularx}{0.47\textwidth}{r S[table-format=4.2] *{4}{S[table-format=3.2]}  }
\toprule
$m_{\text{dm}}$ & \multicolumn{5}{c}{$\langle\sigma_{\text{A}}v\rangle [10^{-23}\text{cm}^3\text{s}^{-1}]$ for NFW profile} \\
\cmidrule(r){2-6} 
[GeV] & $b\bar{b}$ &  $W^{+}W^{-}$ &  $\mu^{+}\mu^{-}$ &  $\tau^{+}\tau^{-}$ &  $\nu\bar{\nu}$  \\
\midrule
10 &  {53.4$\cdot 10^3$} &  \textemdash &  25.1 &  33.4 &  1.46 \\
20 &  269 &  \textemdash &  3.43 &  4.25 &  0.40 \\
30 &  89.1 &  \textemdash &  1.75 &  2.10 &  0.32 \\
40 &  56.9 &  \textemdash &  1.39 &  1.69 &  0.33 \\
50 &  38.7 &  \textemdash &  1.22 &  1.46 &  0.25 \\
100 &  20.6 &  3.29 &  1.03 &  1.18 &  0.42 \\
200 &  16.2 &  4.49 &  1.44 &  1.53 &  0.87 \\
300 &  15.7 &  5.89 &  2.13 &  2.18 &  1.86 \\
400 &  16.4 &  7.28 &  2.94 &  2.84 &  2.88 \\
500 &  17.3 &  8.40 &  3.71 &  3.37 &  4.38 \\
1000 &  22.8 &  14.7 &  9.57 &  7.66 &  26.2 \\
\bottomrule
\end{tabularx}
\end{table}

\begin{table}
\caption{Upper limits on the self\hyp{}annihilation cross section assuming the Burkert halo profile.}
\label{tab:nfw}
\begin{tabularx}{0.47\textwidth}{r S[table-format=4.2] *{1}{S[table-format=3.2]}  *{3}{S[table-format=3.2]} }
\toprule
$m_{\text{dm}}$ & \multicolumn{5}{c}{$\langle\sigma_{\text{A}}v\rangle [10^{-23}\text{cm}^3\text{s}^{-1}]$ for Burkert profile} \\
\cmidrule(r){2-6} 
[GeV] & $b\bar{b}$ &  $W^{+}W^{-}$ &  $\mu^{+}\mu^{-}$ &  $\tau^{+}\tau^{-}$ &  $\nu\bar{\nu}$  \\
\midrule
10 &  {132$\cdot 10^3$} &   \textemdash &  47.12 &  64.35 &  3.22 \\
20 &  578 &   \textemdash &  9.67 &  12.9 &  1.35 \\
30 &  230 &   \textemdash &  5.81 &  7.47 &  1.16 \\
40 &  164 &   \textemdash &  4.88 &  6.17 &  1.35 \\
50 &  119 &   \textemdash &  4.50 &  5.75 &  1.31 \\
100 &  74.2 &  15.6 &  4.96 &  5.92 &  2.15 \\
200 &  67.3 &  22.7 &  7.39 &  8.04 &  4.79 \\
300 &  69.9 &  29.3 &  10.7 &  11.2 &  8.41 \\
400 &  73.3 &  35.8 &  14.8 &  14.5 &  14.9 \\
500 &  79.7 &  42.5 &  19.2 &  18.1 &  24.5 \\
1000 &  110 &  76.3 &  52.3 &  42.4 &  187 \\
\bottomrule
\end{tabularx}
\end{table}

\begin{figure}
  \includegraphics[width=\figwidth]{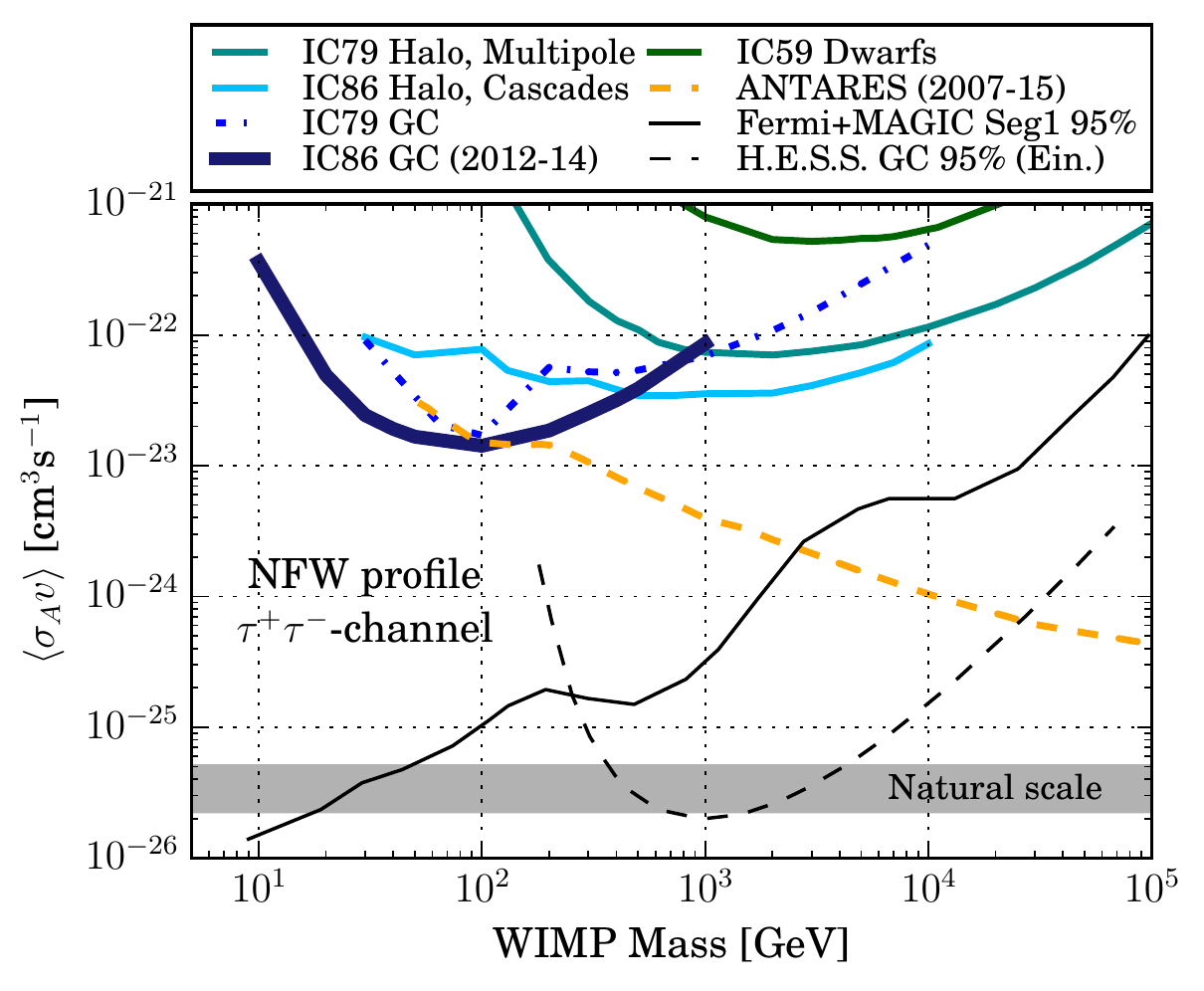}
\caption{Comparison of upper limits on $\langle\sigma_{\text{A}}v\rangle$ versus WIMP mass, for dark matter self\hyp{}annihilating through $\tau^+\tau^-$ to neutrinos, assuming the NFW profile. 
This work (IC86 (2012-14)) is compared to other published searches from IceCube \cite{Aartsen:2013ex,Aartsen:2015hu,Collaboration:2016vk,Aartsen:2015fa} and ANTARES \cite{dmantares}. 
Also shown are upper limits from gamma-ray searches from the dwarf galaxy Segue 1 (Seg1) by FermiLAT+MAGIC \cite{MAGICcollaboration:2016el} and from the galactic center by H.E.S.S. \cite{Abdallah:2016cd}. The `natural scale' refers to the value of $\langle\sigma_{\text{A}}v\rangle$ that is needed for WIMPs to be a thermal relic \cite{Steigman:2012hx}.}
\label{fig:finallimit}
\end{figure}

IceCube has previously searched for a neutrino signal from annihilating dark matter in the center of the Milky Way, using a combined event selection at low and high energies. 
The low energy selection observed an underfluctuation that resulted in an enhanced limit on $\langle\sigma_{\text{A}}v\rangle$, while the high energy selection gave access to higher energies.
This analysis improves on the previous result at most of the energies considered.
In order to compare this work to previous results, Figure \ref{fig:finallimit} shows the upper limits on $\langle\sigma_{\text{A}}v\rangle$ for the $\tau^+\tau^-$ annihilation channel and NFW halo profile of this work to previous results from IceCube and other indirect dark matter detection experiments.
It can be seen that the analysis presented in this paper sets the best limits of a neutrino experiment on WIMP self-annihilation in the galactic center for WIMPs with masses between 10 and 100\,GeV annihilating to $\tau^+\tau^-$.

\section{Conclusions}
\label{sec:conc}
This analysis demonstrates the continued improvements in dark matter searches with neutrinos, providing a valuable complement to the bounds from Cherenkov telescopes and gamma-ray satellites.
A more inclusive event selection and the use of an improved event reconstruction algorithm have increased the sensitivity of IceCube to the signal of dark matter self\hyp{}annihilation.
However, no significant excess above the expected background has been observed in 3 years of Icecube/DeepCore data. 
Upper limits have been put on $\langle\sigma_{A}v\rangle$ providing the leading limits on WIMPs with a mass between 10-100\,GeV for a neutrino observatory.

\begin{acknowledgements}
We acknowledge the support from the following agencies:
U.S. National Science Foundation-Office of Polar Programs,
U.S. National Science Foundation-Physics Division,
University of Wisconsin Alumni Research Foundation,
the Grid Laboratory Of Wisconsin (GLOW) grid infrastructure at the University of Wisconsin - Madison, the Open Science Grid (OSG) grid infrastructure;
U.S. Department of Energy, and National Energy Research Scientific Computing Center,
the Louisiana Optical Network Initiative (LONI) grid computing resources;
Natural Sciences and Engineering Research Council of Canada,
WestGrid and Compute/Calcul Canada;
Swedish Research Council,
Swedish Polar Research Secretariat,
Swedish National Infrastructure for Computing (SNIC),
and Knut and Alice Wallenberg Foundation, Sweden;
German Ministry for Education and Research (BMBF),
Deutsche Forschungsgemeinschaft (DFG),
Helmholtz Alliance for Astroparticle Physics (HAP),
Initiative and Networking Fund of the Helmholtz Association,
Germany;
Fund for Scientific Research (FNRS-FWO),
FWO Odysseus programme,
Flanders Institute to encourage scientific and technological research in industry (IWT),
Belgian Federal Science Policy Office (Belspo);
Marsden Fund, New Zealand;
Australian Research Council;
Japan Society for Promotion of Science (JSPS);
the Swiss National Science Foundation (SNSF), Switzerland;
National Research Foundation of Korea (NRF);
Villum Fonden, Danish National Research Foundation (DNRF), Denmark
\end{acknowledgements}



\end{document}